\documentclass{aa}
\usepackage{graphicx}
\usepackage{txfonts}
\def\gapprox{{_>\atop{^\sim}}}
\def\lapprox{{_<\atop{^\sim}}}

\def\cmmd{\rm {cm^{-3}}}

\def\s-1{\rm {s^{-1}}}
\def\twco{$^{12}$CO}

\def\etal {et al.}
\def\kms {\hbox{${\rm km\,s}^{-1}$}}

\begin{document}
   \title{Luminous HC$_3$N line emission in NGC~4418}

   \subtitle{Buried AGN or nascent starburst?}

   \author{S. Aalto\inst{1}
          \and
          R. Monje\inst{1}
        \and
        S. Mart\'{\i}n\inst{2}
          }

   \offprints{S. Aalto}

   \institute{Onsala Space Observatory,
              S-439 94 Onsala, Sweden\\
              \email{raquel@oso.chalmers.se}
        \and
        Harvard-Smithsonian Center for Astrophysics, 60
Garden Street, Cambridge, MA 02138, USA
             }

   \date{received, accepted}

  \abstract
  {}
   {To investigate the properties of the nuclear molecular gas and to address
the nature of the deeply buried source driving the IR emission of NGC~4418. }
   {IRAM 30m observations and basic non-LTE, single component radiative transport modelling of HNC, HCN,
   HCO$^+$, CN, HC$_3$N and H$_2$CO}
   {We find that NGC~4418 has a rich molecular chemistry - including unusually luminous HC$_3$N $J$=10--9, 16--15
and 25--24 line emission - compared to the other high density tracers.
We furthermore detect: ortho-H$_2$CO 2--1, 3--2; CN 1--0, 2--1; HCO$^+$, 1--0. 3--2,
HCN 3--2, HNC 1--0, 3--2 and tentatively OCS 12--11.
The HCN, HCO$^+$, H$_2$CO and CN line emission can be fitted to
densities of $n=5 \times 10^4 - 10^5$ $\cmmd$ and gas temperatures $T_{\rm k}$=80--150 K.
Both HNC and HC$_3$N are, however, significantly more
excited than the other species which requires higher gas densities - or radiative excitation through
mid-IR pumping. 
The HCN line intensity is fainter than that of HCO$^+$ and HNC for the 3--2 transition,
in contrast to previous findings for the 1--0 lines where the HCN emission is the most luminous.
Assuming all line emission is emerging from the same gas, abundances of the observed species are
estimated to be similar to each other within
factors of 2-5. The most noteworthy being a high abundance of HC$_3$N and a small-to-moderate
abundance ratio between HCN and HCO$^+$.  
}
   {We tentatively suggest that the observed molecular line emission is consistent with 
   a young starburst, where the emission can be understood as emerging from dense, warm gas
   with an additional PDR component. We find that X-ray chemistry is not required to explain the 
   observed mm line emission, including the HCN/HCO$^+$ 1--0 and 3--2
   line ratios.
   The luminous HC$_3$N line emission is an expected signature of dense, starforming gas.
   A deeply buried AGN can not be excluded, but its impact on the surrounding molecular
   medium is then suggested to be limited. However, detailed
   modelling of HC$_3$N abundances in X-ray dominated regions (XDRs) should be carried out. The possibility
   of radiative excitation should also be further investigated.
}

%
\maketitle

%

\section{Introduction}

Since their discovery by the Infrared Astronomical Satellite (IRAS), 
luminous infrared galaxies (LIRGs, $L_{\rm IR}> 10^{11}$ L$_\odot$)
have been studied at almost all wavelengths (e.g. Sanders \& Mirabel
1996). They radiate most of their luminosity as dust thermal
emission in the infrared. However, the nature of the main power
source (a starburst, an active galactic nucleus (AGN) or a
combination of both) is still unclear.

\begin{table*}
\caption{\label{beam} Observational parameters and Gaussian line fits.}
\begin{tabular}{lcccccccc}
Line & $\nu$$^{\rm a}$ [GHz]  & HPBW$^{\rm b}$ [$''$] & $\eta_{\rm mb}$$^{\rm c}$ &  $I$$^{\rm d}=\int T_{\rm A}^*
dV$ [K \kms]
&
$V_{\rm c}$$^{\rm e}$ [\kms] & $\Delta V$$^{\rm f}$ [\kms] & $T_{\rm A}^*$$^{\rm h}$ [mK]\\[2pt] 
\hline
\hline \\
CO  1--0 & 115.271 & 21  & 0.74 & $14.6 \pm 0.7$ & 2110 & 120 & 114.0  \\
CO  2--1 & 230.538 & 10  & 0.53 & $20.2 \pm 0.9$ & 2114 & 107 & 157.0  \\
CN  1--0 & 113.491 & 22  & 0.74 & $1.50 \pm 0.25$ & 2081 & 154 & 8.5  \\
CN  2--1$^{\rm i}$ & 226.875 & 11  & 0.53 & $2.46 \pm 0.2^{\rm j}$ & 2115 & 150 & 7.8 \\
CN  2--1$^{\rm k}$ & 226.659 & 11  & 0.53 &  \dots                 & \dots & 215 & 5.5  \\
HCO$^+$  1--0 & 89.188 & 28  & 0.77 & $1.74 \pm 0.12$ & 2110 & 150 & 9.9 \\
HCO$^+$ 3--2 & 267.558 & 9.2  & 0.45 & $4.1 \pm 0.6$ & 2120 & 190 & 21.5  \\
HCN 3--2 & 265.886 & 9.4  & 0.46 & $2.4 \pm 0.3$ & 2110 & 240 & 9.3  \\
HNC  1--0 & 90.664 & 27  & 0.77 &  $1.24 \pm 0.12$ & 2120 & 156 & 7.5 \\
HNC 3--2 & 271.981 & 9  & 0.44 & $4.97 \pm 0.6$ & 2120 & 150 & 29.3  \\
HC$_3$N  10--9 & 90.979 & 27  & 0.77 & $0.8 \pm 0.08$ & 2110 & 122 & 6.4 \\
HC$_3$N  16--15$^{\rm l}$ & 145.561 & 17  & 0.70 & $1.7 \pm 0.08$ & 2120 & 130 & 8.6 \\
HC$_3$N  25--24 & 227.419 & 11  & 0.53 & $1.6 \pm 0.2$ & 2130 & 140 & 10.0  \\
HNCO 4--3 & 87.925 & 28  & 0.77 & $<$ 0.36 (3$\sigma$) & \dots & \dots & \dots\\
H$_2$CO 2--1 & 140.839 & 17  & 0.70 & $1.05 \pm 0.2$ & 2156 & 145 & 5.7  \\
H$_2$CO 3--2 & 225.698 & 11  & 0.53 & $0.6 \pm 0.3$ & 2140 & 145 & 3.6 \\
OCS 12--11 & 145.947 & 17 & 0.70 &  $0.8 \pm 0.14$ & \dots & 174 & 4.8  \\

\hline \\
\end{tabular} \\
a) Rest frequency, b) Beam width, c) Main beam efficiency  d) Integrated intensity 
in $T_{\rm A}^*$, uncorrected for beam efficiency and beam size. 
($T_{\rm A}^*$ is transferred into
the main beam brightness scale, $T_{\rm mb}$, through multiplying $T_{\rm A}^*$ with the
beam efficiency $\eta_{\rm mb}$.) For the 30m telescope, the relation between $T_{\rm mb}$
and flux is $S$=4.95 Jy/K. e) Center velocity of the line,
f) Gaussian fit line width, h) Line fit peak intensity.\\
i) First spingroup, j) Including both first and second spingroup. 
k) Second spingroup, l) HC$_3$N  16--15 is contaminated by para-H$_2$CO. Its contribution is
estimated to 20\%. Fitted Gaussians
and integrated intensities are only for the HC$_3$N component (see text).
\end{table*}

NGC~4418 is a nearly edge-on Sa-type galaxy with deep mid-infrared
silicate absorption features suggesting that the inner
region is enshrouded by large masses of warm (85 K) dust (e.g. Spoon \etal\
2001, Evans \etal\ 2003). Due to the high
obscuration, it is difficult to determine the
nature of the activity that is driving the luminosity. The
infrared luminosity to molecular gas mass ratio - $L_{\rm IR}$ /
M(H$_2$) = 100 is high for a non-ULIRG galaxy (Sanders, Scoville, \&
Soifer 1991). Furthermore, NGC~4418 has a high infrared to radio continuum ratio
$q\approx 3$ ($q$= the logarithmic ratio
of far-infrared (FIR) to radio flux densities). This FIR-excess may be
caused by a young, somewhat synchrotron deficient, starburst (Roussel \etal\ 2003).
The lack of hard X-ray emission (Cagnoni \etal\ 1998) and a resolved radio continuum emission
support the starburst interpretation. 

In contrast, Seyfert-like infrared colors indicate that the dust in
NGC~4418 is being heated by an obscured AGN. Imanishi \etal\ (2004)
find broad NIR H$_2$ lines and a mm HCN/HCO$^+$ $J$=1--0 line ratio
of $>$1  --  which, they suggest, support the notion of an obscured AGN.
The lack of hard X-rays may in this context be due to the AGN being compton thick.

The molecular line ratio HCN/HCO$^+$ 1--0 has been suggested
as a diagnostic tool for identifying the effect of an AGN on its surrounding
molecular ISM (e.g. Kohno \etal\ 2001, Imanishi \etal\ 2004). The X-rays
from the AGN impact the interstellar medium (ISM) surrounding it,
creating an XDR -- an X-ray Dominated Region. Since X-rays may penetrate 
deep into the surrounding ISM, the resulting XDR may potentially become large
and affect molecular properties on large scales.
Molecular line ratios are therefore potentially useful as XDR indicators. 
According to models by e.g. Maloney \etal\ (1996), 
HCO$^+$ is destroyed in XDRs  -- while HCN may enjoy high abundances, resulting
in an elevated  HCN/HCO$^+$ 1--0 line ratio. Recent XDR models, however, question
the suggested underabundance of HCO$^+$ in XDRs (e.g. Meijerink and Spaans
(2005), Meijerink \etal\ 2006, 2007) and find HCN and HCO$^+$ abundances to be similar -- or
HCO$^+$ to be somewhat overabundant.  Thus, the interpretation
of an elevated HCN/HCO$^+$ 1--0 line ratio is not straight forward, in particular
when line opacities and excitation are taken into account.
Furthermore, dense gas with low ionization levels may also have a relative
underabundance of HCO$^+$. Therefore, further observations and modelling is necessary
to asses the impact of AGNs and starbursts on their surrounding ISM.

We have carried out an initial molecular line search of NGC~4418
with the IRAM 30 telescope. The goal was to observe multiple transitions
of high density tracer molecules and to search for complex molecules
to get a first handle on the molecular ISM conditions of NGC~4418. 
High density tracer molecules require gas densities in excess
of $10^4$ $\cmmd$ in order to be collisionally excited. This is two orders of
magnitude greater than the \twco\ 1--0 line which is the most commonly used 
tracer of molecular gas in external galaxies. 

We find a surprisingly rich ISM chemistry in NGC~4418 -- with very luminous HC$_3$N
line emission - and that the HCN 3--2 line emission is fainter than both HCO$^+$ and HNC 3--2.
In section 2, observations and results in terms of line intensities,
ratios, and RADEX models are presented. In section 3, we discuss the interpretation of 
the luminous HC$_3$N line and the other line intensities in the context of 
XDR and starburst scenarios. We tentatively suggest a starburst scenario and
briefly present how this conclusion may be tested.

\section{Observations and results}

We have used the IRAM 30m telescope to observe 1mm, 2mm and 3mm lines of HNC, HCO$^+$, HCN,
CN, HC$_3$N, Ortho-H$_2$CO, CO and HNCO towards the center of the luminous galaxy NGC~4418. 
Observations were made in May and July 2006, and the system noise temperatures were
typically 150 K and 500 K for the 3 mm and 1 mm lines, respectively.
Pointing was checked regularly on nearby continuum sources and rms was found
to vary between 1.5$\arcsec$ and 2$\arcsec$. Beamsizes and efficiencies are
shown in Table~1.

First order baselines have been removed from all spectra. 
Weather conditions during both runs were close to normal summer
conditions with precipitable water vapour, pwv = 4.3--5.1~mm in May
and pwv = 3.5--4.1~mm in June. 
The 1mm observations of HCN and HNC 3--2 were repeated in 
the July run to confirm the line ratios. We found that line intensities 
could be repeated within 15\% suggesting that, in general, 1~mm conditions both
in May and July were stable. We also repeated the H$_2$CO 2--1
2mm line and we found that the line
intensity was repeatable within 10\%.

\begin{figure*}
\resizebox{17cm}{!}{\includegraphics[angle=0]{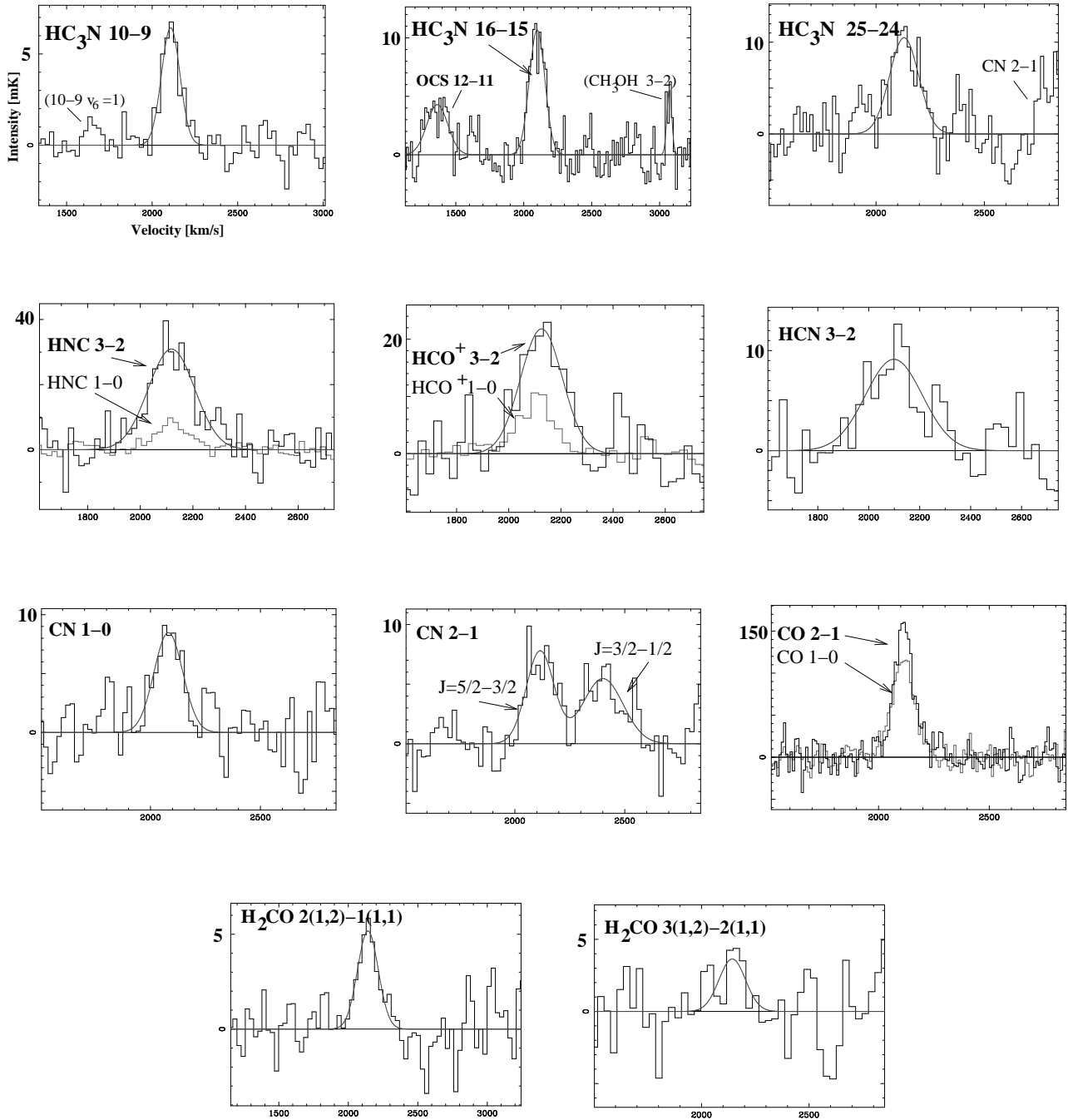}}
\caption{\label{hnc32_n4418} Spectra of detected species towards NGC~4418 ($\alpha$: 12:26:54.63,
$\delta$: -00:52:39.6)(J2000). Note that the intensity
scale is $T_{\rm A}^*$ in mK, and that line intensities shown here are not corrected for beam size. 
Top three panels: Velocity resolution
is 15 \kms. Left: HC$_3$N 10--9 (position for 10--9 $v_6$=1--0 rotational-vibrational transition
indicated).
Center: HC$_3$N 16--15 - somewhat blended with para-H$_2$CO (see text). Tentative line emission from OCS
12--11 towards the band edge (position of CH$_3$OH 3--2 indicated in spectrum).
Right: HC$_3$N 25--24 (at the edge of the band the beginning of the CN 2--1 line (CN line
intensity cannot be trusted).  
Second row of panels: Velocity resolution
is 30 \kms. Left: HNC 3--2 and 1--0, Center: HCO$^+$ 3--2 and 1--0, Right: HCN 3--2.
Third row of panels: Left: CN 2--1 (note the two spin groups), Center: CN 1--0, Right: CO 1--0 and 2--1 .
Last row of two panels: Left: Ortho H$_2$CO 2--1, Right: Ortho H$_2$CO 3--2.  }
\end{figure*}

\begin{table}[htb!]
\caption{\label{beam} Observed Line Ratios$^{\rm a}$}
\begin{tabular}{lccc}
Species & Transitions & Line ratio\\[4pt]
\hline
\hline \\

CO & 2--1/1--0 & 0.50\\
HCO$^+$ & 3--2/1--0 & 0.45\\
HNC & 3--2/1--0 & 0.80\\
HCN & 3--2/1--0 & 0.20$^{\rm b}$\\
CN & 2--1/1--0 & 0.40$^{\rm c}$\\
H$_2$CO & 3(1,2)--2(1,1)/2(1,2)--1(1,1) & 0.35\\
HC$_3$N & 16--15/10--9 & 0.90$^{\rm d}$\\
HC$_3$N & 25--24/10--9 & 0.50\\

\hline \\
\end{tabular} \\
a)If $\theta_{\rm s}$ is the gaussian source size (which we have assumed to be 2$\arcsec$ (see text)),
and $\theta_{\rm b}$ is the beam size for the line,
the ratio of integrated intensities for transition $X$ and $Y$, corrected for beam and
source-size, is assumed to scale according to: $ { \int T_{\rm mb}(X) dV \over \int T_{\rm mb}(Y) dV} = 
 { \int T_{\rm A}^*(X) dV \, (\theta_{\rm b}^2(X) + \theta_{\rm s}^2) \, \eta_{\rm mb}(X) \over
 \int T_{\rm A}^*(Y) dV \, (\theta_{\rm b}^2(Y) + \theta_{\rm s}^2) \, \eta_{\rm mb}(Y)}$ 
Errors in line ratios are $\approx$20\% including rms
and calibration errors. 

b): Peak to peak ratio is 0.1 - but integrated 
intensity ratio is 0.2. HCN 1--0 data from Imanishi \etal\ (2004). \\
c): The integrated line ratio should
include all spingroups for 1--0 and 2--1. We only have the brightest of the two main 1--0 spingroup
in our beam since the other spingroup occurs +800 \kms away from the main.  Integrated intensity
ratio is 0.4 - peak-to-peak is 0.3. d) Line ratio somewhat uncertain due to para-H$_2$CO
contamination (see text).

\end{table}

\subsection{Line intensities and ratios}

We have detected the following line emission:\, HNC 1--0, 3--2; HCO$^+$ 1--0, 3--2;
HCN 3--2, CN 1--0, 2--1; HC$_3$N 10--9, 16--15,
25--24; Ortho-H$_2$CO 2--1 and 3--2; CO 1--0, 2--1 (see Fig. 1, Table~1). We also tentatively
detect the OCS 12--11 line at the edge of the HC$_3$N 16--15 band. HNCO 4--3 was not detected
down to an rms level of 2~mK. In Fig.~1 we have marked the position of the HC$_3$N $J$=10--9, v$_6$=1--0
rotational-vibrational transition, and the CH$_3$OH 3--2 line. They are both 2$\sigma$ detections
if one considers the rms only. Further observations are required to confirm their presence.
Integrated line intensities and fitted gaussian line parameters are presented in Table~1.
A para-H$_2$CO 2--1 line is blended with the HC$_3$N 16--15 line (para-H$_2$CO is shifted 
+42 MHz, or -82 \kms from the HC$_3$N line). A gaussian fit where we set the velocity
of the line center to 2120 \kms and the line width to 130 \kms result in an estimate of the 
para-H$_2$CO contribution to 20\% of the total integrated line intensity. This implies an H$_2$CO
2--1 ortho/para line ratio of $>$2 (see Table~1 for the Ortho 2--1 line intensity). In warm
gas ($T_{\rm k} \gapprox 40$K) the ortho-to-para ratio is close to the statistical ratio of 3
(e.g. Kahane \etal\ 1984).

Line ratios, corrected for beam size and efficiency, are presented in Table~2. When correcting
for beam size we require an estimate on the source size $\theta_{\rm s}$. 
From Imanishi \etal\ (2004) one can deduce an
upper limit to the HCN 1--0 source size of 2$\arcsec$. We adopt this as 
$\theta_{\rm s}$ for our calculations -- assuming that all high density tracer emission
is emerging from the same region. Errors in line ratios are generally 20\% including rms
and calibration errors, but not including source size errors.

\subsection{Radiative transport modelling}

We used the online mean escape probability (MEP) code RADEX
(Sch\"oier \etal\ 2005) for radiative transport modelling of the
observed molecular line ratios. 

The online version of RADEX assumes spherical
cloud geometry. We assume
that all line emission is emerging from the same gas.  
This is of course unlikely to be true, but will give a first
handle on physical conditions and relative column densities.

RADEX models the average condition for a single cloud type.
Here, the emission has been modelled as emerging from one structure of
linewidth 150 \kms, dealing with the global line emission at once. 
A more realistic approach is to assume an ensemble
of individual clouds with a filling factor. Furthermore, multiple cloud types,
and/or gradients within clouds should also be considered in a more advanced model - 
including issues on self-gravitation, alternative excitation and dynamics.
This will be presented in paper II, where we also will 
investigate the radiative excitation further. The aim of the simple approach here
is to look at average relative abundances, and physical conditions indicated by the
line ratios.

\subsubsection{HCN, HCO$^+$, H$_2$CO and CN}

The dust temperature of NGC~4418 is estimated to be high, $T_{\rm d}$=85 K - and emerging
from a compact nuclear structure of diameter $\approx 70$ pc (Evans \etal\ 2003).
The dense molecular gas appears also to be compact and at
number densities exceeding $10^4$ $\cmmd$, the gas temperature should begin to approach
that of the dust. Thus assuming that the gas is also warm, we have varied the gas kinetic temperature
from $T_{\rm k}$=50 to 150 K in our RADEX models.
We have varied the gas densities $n$ within the range $10^4 - 10^6$ $\cmmd$. Lower densities should
not result in significant line emission from any of the observed high density tracer species -- unless
affected by radiative excitation (see below). Higher densities are unlikely due to the moderate excitation.

We searched for a solution that would both accomodate the greater HCN 1--0 luminosity
than that of HCO$^+$ 1--0 (Imanishi \etal\ 2004) -- and the reverse situation for the 3--2 line, and at the same
time would fit the low excitation of HCN (and the higher excitation of HCO$^+$).
Solutions within the constraints given above could be found for a density range of
$n=5 \times 10^4 - 10^5$ $\cmmd$. Higher densities are not allowed since HCO$^+$ becomes thermalized.
Lower densities give too low excitation even when the temperatures are allowed to exceed 150 K.
The column density ratio between HCN/HCO$^+$ varies between 3 and 5 for the whole range of
solutions. For $n=5 \times 10^4$ $\cmmd$ the best fit is found for a temperature of 150 K, for
$n=1 \times 10^5$ best fit is for $T_{\rm k}$=80 K.

Ortho-H$_2$CO is also moderately excited, and the 3--2/2--1 line ratio agrees reasonably well with the
above physical conditions, even if temperatures exceeding 80 K and 150 K, respectively, would give a better fit. The
column density ratio between H$_2$CO/HCO$^+$ ranges between 1 and 3 for all solutions.

Online RADEX currently contains no datafiles for handling CN, but
a rough comparison to CS can be made, and run RADEX for CS with the above model parameters
which suggests that the CN line emission may also be fitted to the above conditions.
CN appears to be on average equally abundant as HCO$^+$.\\

The \twco\ 2--1/1--0 line ratio is low (Table~2) and requires densities $n \lapprox 10^3$ $\cmmd$
and may not be fitted to the above model. This is either due to the \twco\ emission emerging
from a low density nuclear component - or the notion of a 2$\arcsec$ source size is not applicable for 
the lower transition \twco\ emission. OVRO observations by Dale \etal\ (2005) of the CO emission show a
structure unresolved in their 5$\arcsec$ beam.

\subsubsection{HNC and HC$_3$N}

The HC$_3$N and HNC line luminosities are surprisingly high --- and
significantly more excited than the other high density species.
This points to either the possibility that HNC and HC$_3$N are affected by radiative
excitation - or that they are emerging from a denser and/or warmer cloud component
than the other species.\\

\noindent
{\it Collisional excitation:} 

\noindent
If we adhere to the same constraints on conditions and search in the same
range as in section 2.2.1., we find that the HC$_3$N $J$=10--9 and 16--15 lines can be
fitted to a component of density $n=2 \times 10^5$ $\cmmd$ and temperature $T_{\rm k}$=80 K. In this
context, the HC$_3$N column density is high, similar to the HCN abundance.
The observed 25--24 line is, however, too bright to
be fitted to this model. More than 90\% of the 25--24 line would have to be emerging from
another component with a density $n > 10^6$ $\cmmd$ to produce a luminous 25--24 line. 
Such a gas component would contaminate also the other high density gas tracer lines -- where the low
excitation does not leave much room for such a very dense gas phase. The filling factor of a
$n=10^6$ $\cmmd$ gas component would also have to be low.

A collisional exitation of HNC would require densities of $n \gapprox 5 \times 10^5$ $\cmmd$ for
the given temperature range - which in the contaxt of a one component model is not an
allowed density range for the HCN, HCO$^+$ and H$_2$CO line ratios. \\

\noindent
{\it Radiative excitation:} 

\noindent
Pumping of HNC by 21.5 $\mu$m radiation via the degenerate
bending mode has been discussed by Aalto \etal\ (2007) as a
possibility to explain the large HNC/HCN 3--2 line ratio in
NGC~4418, Arp~220 and Mrk~231. For HNC, a mid-infrared background with a brightness
temperature of 50 K will start to successfully compete with collisions at gas
densities $n<10^4$ $\cmmd$. Thus, significant contribution to the line
emission may then originate from a gas phase of lower density than 
the critical density for the species.  

\noindent
HC$_3$N may also be pumped via a number of infrared bending transitions, e.g.(Wyrowski \etal\ 1999):
\vspace{-0.2cm}
\begin{itemize}
\item Bending mode v$_5$=1--0 at 663 cm$^{-1}$ (956 K) with $A_{ul}$ = 2.2 s$^{-1}$, 15 $\mu$m (HCC bend)
\item Bending mode v$_6$=1--0 at 498 cm$^{-1}$ (718 K) with $A_{ul}$ = 0.15 s$^{-1}$, 20 $\mu$m (CCN bend)
\item Bending mode v$_7$=1--0 at 223 cm$^{-1}$ (321 K) with $A_{ul}$ = $6\times10^{-4}$ s$^{-1}$, 45 $\mu$m (CCC bend)
\end{itemize} 

\noindent
In Fig.~1 we indicate the position of the $J$=10--9 v$_6$ rotational-vibrational line as a 2$\sigma$ detection, but 
more observations are needed to show whether this feature is real, or just a baseline uncertainty.
Rotational-vibrational transitions of HC$_3$N have been observed in the Galaxy towards hot cores
(e.g. Wyrowski \etal\ 1999). 

\noindent
{\it Column density comparison}

If HC$_3$N line emission is affected by radiative excitation, it may have a larger filling factor than
the line emission from unaffected molecular lines, such as HCN. 
Column density estimates done under the assumption of collisional excitation from the same gas may therefore
overestimate the relative HC$_3$N abundance.

For NGC~4418 this may largely be related to the mass fraction of lower density gas, in relation to the
high density gas fraction. We expect the HC$_3$N abundances to be lower in the low density gas phase since this
phase should be more vulnerable to photodestruction. Furthermore, the mass fraction of low density gas in galactic
nuclei is often found to be significantly lower than that of the high density gas mass fraction, even if the
lower density gas has a higher filling factor (e.g. Aalto \etal\ 1994, 1995). We therefore estimate the errors in
the abundance estimate due to potential radiative excitation, as rather small: within factors of a few. Errors due
to insufficient data, and model oversimplification may be significantly larger. Observations of more HC$_3$N transitions
will help establish 
to what degree radiative excitation may affect the line intensities of the rotational transitions.

\section{Discussion}

\subsection{Interpreting the HC$_3$N line emission}

{\it The relative HC$_3$N 10--9 line emission in NGC~4418 is one of the
highest measured in a galaxy on these scales -- comparable to that
found for Arp~220 }(Aalto \etal\ 2002; Martin-Pintade \etal\ in preparation).
The implied HC$_3$N column densities rival those of HCN
suggesting unusually large abundances.

How can we understand the luminous HC$_3$N line emission? Below we list
three scenarios and their expected impact on HC$_3$N abundances:

\begin{itemize}
\item Warm dense gas: Rodriguez-Franco \etal\ 1998, find large  HC$_3$N/CN abundance ratios in
hot cores (warm, star-forming, dense gas) which they attribute to very large
HC$_3$N abundances (\,$X[{\rm HC}_3{\rm N}] \approx 10^{-8}$) caused largely by evaporation of icy 
grain mantles and low destuctive cosmic ray and UV fluxes. HC$_3$N can also be formed via
the neutral-neutral reaction: 
C$_2$H$_2$ + CN $\rightarrow$ HC$_3$N + H (e.g. Meier and Turner 2005). \\

\item PDRs: HC$_3$N is easily destroyed by UV photons (destructon rate $5.5 \times 10^{-9}$ s$^{-1}$)
and very rapidly via reactions with the ions C$^+$ and He$^+$ (Prasad \& Huntress, 1980;
Turner \etal\ 1998).
Thus, we do not expect luminous HC$_3$N line emission from PDRs since it would quickly
become destroyed by photodissociation or by reaction with C$^+$. This is also observed 
by Rodriguez-Franco and collaborators towards PDRs in the Orion complex.\\

\item XDRs: X-rays can penetrate large columns of gas and are therefore not as good as UV photons at dissociating
molecules. There are UV photons in an XDR, but mostly in the form of relatively
weak, secondary far-UV photons. Their impact on HC$_3$N destruction is likely less
than in a PDR. An XDR is characterized (among other things) by high abundances of C$^+$ and in particular
He$^+$ - that coexist with neutrals like CO and C. Both  C$^+$ and He$^+$ are sources of destruction
of HC$_3$N. The He$^+$ abundances in XDRs can be five times higher than those in PDRs according to
to XDR/PDR models (Meijerink and Spaans 2005). In the absence of icy-mantle grain processing
and of important ion-neutral fomation mechanisms, the neutral-neutral formation mechanism indicated
above will be too slow to keep up with the destruction proceeses. 
\end{itemize}

\noindent
HC$_3$ abundances are low in PDRs and we expect them also to be so in XDRs - although no direct
obervations exist, and detailed modelling is required to determine abundances. From Galactic work, it
is well known that the molecule thrives in hot cores and in cold dark clouds. It has been observed
at high resolution in the nearby galaxy IC~342
by Meier and Turner (2005) where they find the 10--9 emission to be strongly correlated with 3~mm continuum
emission interpreted as sites of on-going starformation.

\subsection{The HCN/HCO$^+$ 1--0 line ratio and XDRs}

HCO$^+$ is suggested by e.g. Kohno \etal\ 2001 to be a tracer of
the fraction of the dense gas which is involved in starformation.
They compared the HCN/HCO$^+$ 1--0 line ratio in a selection
of Seyfert and starburst galaxies and find that
the relative HCO$^+$ 1--0 luminosity is significantly higher
in starbursts. According to models by Maloney \etal\ (1996) a
deficiency of HCO$^+$ is expected near a
hard X-ray source - in an X-ray dominated region (XDR) - resulting 
in an elevated observed HCN/HCO$^+$ line intensity ratio. 
Interestingly, elevated
HCN/HCO$^+$ 1--0 line ratios are found in ULIRGs by Gracia-Carpio (2006) --- and in
general the line ratio
appears to correlate with FIR luminosity. The authors attribute this to the presence of
an AGN where the X-rays affect the chemistry
to impact the HCN/HCO$^+$ abundance ratio. \\

Abundances can not be directly deduced from the HCN/HCO$^+$ 1--0 line ratio - even if
the emission is indeed emerging from the same gas. Excitation effects must be taken into account.
Imanishi \etal\ (2004) report an HCN/HCO$^+$ 1--0 line ratio of 1.8 for NGC~4418, suggesting that
this supports the notion of an XDR. We observe, however, the opposite case for the 3--2 transitions
and we find that both the 1--0 and 3--2 line
ratios can be fitted to a rather moderate HCN/HCO$^+$ abundance ratio (see section 2.2.1).

Furthermore, recent XDR modelling suggest that underabundant HCO$^+$ may not be an expected feature
of an XDR. In contrast, for a large range of conditions, Meijerink and Spaans (2005) and Meijerink
\etal\ (2006) find the HCO$^+$ abundances exceeding those of HCN in the XDR. In their models
HCN is more abundant than HCO$^+$ at the edges of the XDR, but in these regions molecular abundance
are generally low and do not contribute significantly to the total line luminosity. A discussion of
this can also be found in Aalto \etal\ (2007).

Relatively low HCO$^+$ abundances may be expected in young, synchrotron-deficient, starbursts
which would also result in HCN/HCO$^+$ 1--0 line intensity ratios exceeding unity.
An alternative interpretation to an elevated HCN/HCO$^+$ 1--0 line ratio may therefore be
that some Seyfert nuclei are surrounded by regions of very young  starformation.
If the starburst is young enough not to have produced a significant number of
supernovae, then the relative HCO$^+$ 1--0 luminosity may be low compared to
that of HCN 1--0.

\subsection{CN and HNC}

All XDR models predict high
abundances of CN - a radical which is also abundant in PDRs (e.g.
Rodriguez-Franco \etal\ 1998). For both XDRs and PDRs the CN abundance is expected to be significantly
higher (orders of magnitude) higher than that of HCN. For NGC~4418 CN is detected - but line emission
appears to suggest abundances less than HCN.

Luminous HNC emission has been observed towards a sample of warm galaxies
(e.g. Aalto \etal\ 2002) and overluminous HNC 3--2 in three galaxies (Aalto \etal\ 2007).
Low HNC abundances are observed towards hot, dense cores
in the Galaxy (e.g. Schilke \etal\ 1992), while in a PDR, equal HCN and HNC abundances are
expected (e.g. Aalto \etal\ 2002, Meijerink \etal\ 2006). In dense XDRs ($ n \gapprox 10^5$ $\cmmd$)
Meijerink \etal\ 2006 predict HNC/HCN abundance ratios of two. Mid-IR pumping may boost the
HNC luminosity relative to the HCN luminosity.
For NGC~4418 the HNC emission may either emerge from a PDR component (and affected by radiative excitation)
or it emerges from a deeply embedded, dense XDR.

\subsection{NGC~4418: Buried AGN or nascent starburst?}

We find a rich molecular chemistry in NGC~4418 and luminous HC$_3$N line emission which in our Galaxy is
associated with warm, dense starforming cores. In this context, the elevated HCN/HCO$^+$ 1--0 line ratio 
found by Imanishi \etal\ (2004) may instead indicate young starformation. 
As discussed in section 3.2, we therefore advice caution in using the HCN/HCO$^+$ 1--0 line ratio as a
diagnostic tool for XDR activity since its interpretation is ambiguous. It is clear that further XDR modelling
is required, including a continued dicussion of the HCN/HCO$^+$ line emission ratio, as well as modelling
of HC$_3$N abundances in XDRs.

For NGC~4418, we tentatively suggest that the line emission can be fitted to a combination of a warm,
dense gas component not strongly affected by UV or cosmic rays (hot-core like
chemistry) supplying abundant HC$_3$N and HCN --- with an additional PDR component (providing CN, HCO$^+$ and
HNC line emission). This scenario is consistent with the suggestion by Roussel \etal\ that
the high $q$-factor is due to a young starburst. They also propose that NGC~4418
is probably in a somewhat later evolutionary stage than other high-$q$ objects since it does contain
a synchrotron source, and contains both a nascent starburst and a more evolved component.

\subsubsection{Future observational tests}

Even though we suggest a starburst explanation for the observed line ratios,
a deeply embedded AGN, surrounded by starformation, cannot be excluded. Without
high resolution information we cannot tell
whether the HC$_3$N line emission is emerging from further out than, for instance,
the HCN or HCO$^+$ line emission. In general, observations at higher
resolution will help resolve the dichotomy around XDRs vs. starformation. Below we list
further observational tests of the molecular ISM: 

\begin{itemize}
\item  Alternative, and more unambigious, XDR diagnostics are provided by highly excited
($J>10$) rotational
lines of \twco\ (Meijerink \etal\ 2006, Aalto \etal\ 2007). XDRs produce very warm \twco\
compared to, for example, PDRs. Other lines that can help distinguish between XDRs,PDRs
and young starformation include HOC$^+$, NO, CO$^+$ and CS (e.g. Meijerink \etal\ 2007) 

\item Observe higher $J$ lines of all detected molecular species. This allows for more
detailed modelling of the physical conditions in the gas. For HC$_3$N these
observations --  as well as a search for
the rotational-vibrational lines -- should reveal a possible pumping situation. This is
also important for the HNC line emission.

\item High resolution observations of molecular lines to determine from where the line emission
is emerging. The upcoming ALMA (Atacama Large Millimeter Array) observatory will provide unprecedented
spatial resolution which will allow cloud-scale observations of many of the luminous galaxies.
Dynamical studies of the nuclear gas will enable measurements of enclosed masses, and searches
for molecular outflows.

\end{itemize}

\section{Conclusions}

   \begin{enumerate}
   
    \item We detect line emission from HNC, HCN, HCO$+$ $J$=1--0 and 3--2, 
    CN 1--0 and 2--1, HC$_3$N 10--9, 16--15, 25--24, Ortho-H$_2$CO 3--2, 2--1
   (and tentatively OCS 12 --11) towards the luminous, deeply obscured galaxy NGC~4418.
    All species have low excitation with the exception of HNC and HC$_3$N
    which are more highly excited.

    \item The HC$_3$N line emission is unusually luminous with the 10--9 line having similar 
    intensity as CN 1--0. This suggests large abundances of HC$_3$N, a molecule usually
     associated with young starformation.
      
    \item We find overluminous HNC 3--2 emission - where the HNC
    luminosity is a factor of 2.3 times that of HCN 3--2. This property
    led us to propose that HNC may be pumped by 20 $\mu$m IR continuum
    (Aalto \etal\ 2007). The high excitation of HC$_3$N is also consistent
    with radiative excitation. 

    \item We find an HCN/HCO$^+$ 3--2 ratio of about 0.5. Imanishi \etal\ (2004) report
    the opposite for the 1--0 ratio. The HCN, HCO$^+$, H$_2$CO line ratios
    can be fitted to a single cloud component with $T_{\rm k}$=80 - 150 K and gas densities
    of $5\times 10^4 - 1 \times 10^5$ $\cmmd$ - and an HCN/HCO$^+$ abundance ratio of $\approx$3. 

    \item The HCN/HCO$^+$ 1--0 line ratio has previously been suggested to indicate the presence of
    an XDR associated with an AGN. We find, however, that the molecular
line ratios are also consistent with a young starburst -- in particular the luminous HC$_3$N line emission.
We advice caution in using elevated HCN/HCO$^+$ line ratios as XDR indicators since the interpretation
is not unique.

   \end{enumerate}

\begin{acknowledgements}
      Many thanks to the IRAM staff for friendly help and support. We are grateful to M. Wiedner
      for discussions and comments on the manuscript. 
\end{acknowledgements}


\begin{thebibliography}{}

\bibitem{}
Aalto S., Polatidis A.G., H\"uttemeister S., Curran S.J., 2002, A\&A 381, 783  


\bibitem{}
Aalto, S., Spaans, M.,Wiedner, M. C., H\"uttemeister S., 2007, A\&A, in press


\bibitem{}
Cagnoni, I, della Ceca, R., Maccacaro, T., 1998, ApJ, 493, 54

\bibitem{}
Dale, D. A., Sheth, K., Helou, G., Regan, M. W., H\"uttemeister, S., 2005, AJ, 129, 2197

\bibitem{}
Evans, A. S., Becklin, E. E., Scoville, N. Z., Neugebauer, G.,
Soifer, B. T., Matthews, K., Ressler, M., Werner, M., Rieke, M.,
2003, AJ, 125, 2341

\bibitem{gra06}
Graci\'a-Carpio, J., Garc\'ia-Burillo, S., Planesas, P., Colina, L, 2006, ApJ, 640, 135



\bibitem{}
Imanishi, M., Nakanishi, K., Kuno, N., Kohno, K., 2004, AJ, 128,
2037

\bibitem{}
Irvine, W., M., Goldsmith, P. F., Hjalmarson, \AA, 1987, in {\it Interstellar Processes},
eds. Hollenback and Thronson, Kluwer:Dordrecht, 561 


\item{}
Kahane, C., Lucas, R., Frerking, M. A., Langer, W. D., Encrenaz, P., 1984, A\&A, 137, 211

\bibitem{}
Kohno, K., Matsushita, S., Vila-Vilaro, B., Okumura, S. K., Shibatsuka, T.,
Okiura, M., Ishizuki, S., Kawabe, R., 2001, in {\it The Central Kiloparsec of Starbursts
and AGN: The La Palma Connection}, ASP Conference Series, Vol. 249, eds. Knapen, Beckman,
Shlosman, Mahoney, 672


\bibitem{}
Maloney, P. R., Hollenbach, D. J., Tielens, A. G. G. M., 1996, ApJ,
466, 561

\bibitem{} 
Meier, David S., Turner, Jean L, 2005, ApJ, 618, 259

\bibitem{}
Meijerink, R., Spaans, M., Israel, F.P., 2006, ApJ, 650, L103


\bibitem
Meijerink, R., Spaans, M., Israel, F.P., 2007, A\&A, in press, astro-ph/0610360

\bibitem{}
Prasad, S. S., Huntress, W. T. 1980, ApJ, 239, 151

\bibitem{} 
Rodriguez-Franco, A.,  Martin-Pintado, J., \& Fuente, A., 1998, A\&A, 329, 1097

\bibitem{} 
Roussel, H., Helou, G., Beck, R., \etal, 2003, ApJ, 593,733.

\bibitem{}
Schilke P., Walmsley C.M., Pineau de For\^ets G., \etal, 1992, A\&A 256, 595

\bibitem{} 
Sch\"oier, F.L., van der Tak, F.F.S., van Dishoeck E.F., 
Black, J.H., 2005, A\&A 432, 369

\bibitem{} 
Sanders, D. B. \& Mirabel, I. F. 1996, ARA\&A, 34, 749

\bibitem{} 
Spoon, H. W. W., Keane, J. V., Tielens, A. G. G. M., Lutz, D., Moorwood,
A. F. M., 2001, A\&A, 365, 353

\bibitem{}
Turner, B. E., Lee, H.-H.,  Herbst, E. 1998, ApJS, 115, 91

\bibitem{}
Wyrowski, F., Schilke, P., Walmsley, C. M., 1999, A\&A, 341, 882


\end{thebibliography}
\end{document}